\documentclass[10pt,aps,twocolumn,prc,superscriptaddress,showpacs,nofootinbib,noshowkeys,floatfix,preprintnumbers]{revtex4-2}
\usepackage[dvipdfmx]{graphicx}
\usepackage{graphicx} 
\usepackage[usenames]{color}
\usepackage{amsmath,amssymb}
\usepackage{multirow}
\usepackage{longtable}
\usepackage[normalem]{ulem}
\usepackage{epstopdf}
\usepackage{times}
\usepackage[normalem]{ulem}  % \sout{old text} for strikeout
\allowdisplaybreaks[4]

\renewcommand\sout{\bgroup \color{red} \ULdepth=-.5ex \ULset}
\newcommand{\physdim}[1]{\hspace{1ex} \mathrm{#1}}

\usepackage{bm}

\graphicspath{{./figure/}}
\newcommand{\deceased}[1]{\altaffiliation{#1}}

\begin{document}
\title{Theoretical study of the $\Xi\alpha$ correlation function}
\date{\today}
%\author{Y. Kamiya}
\author{Yuki Kamiya}
\email[]{yuki.kamiya.d3@tohoku.ac.jp}
\affiliation{Helmholtz Institut f\"ur Strahlen- und Kernphysik and Bethe Center for Theoretical Physics, Universit\"at Bonn, D-53115 Bonn, Germany}
\affiliation{RIKEN Interdisciplinary Theoretical and Mathematical Science Program (iTHEMS), Wako 351-0198, Japan}
\affiliation{Department of Physics,
Tohoku University Sendai 980-8578, Japan}

%\author{A. Jinno}
\author{Asanosuke Jinno}
\email[]{jinno.asanosuke.36w@st.kyoto-u.ac.jp}
\affiliation{Department of Physics, Faculty of Science, Kyoto University, Kyoto 606-8502, Japan}

%\author{T. Hyodo}
\author{Tetsuo Hyodo}
\email[]{hyodo@tmu.ac.jp}
\affiliation{Department of Physics, Tokyo Metropolitan University, Hachioji 192-0397, Japan}
\affiliation{RIKEN Interdisciplinary Theoretical and Mathematical Science Program (iTHEMS), Wako 351-0198, Japan}

%\author{A. Ohnishi}
\author{Akira Ohnishi}\deceased{Deceased.}
%\email{ohnishi@yukawa.kyoto-u.ac.jp}
\affiliation{Yukawa Institute for Theoretical Physics, Kyoto University, Kyoto 606-8502, Japan}
%\affiliation{RIKEN Nishina Center, Wako 351-0198, Japan}

\preprint{KUNS-3020}

\begin{abstract}
	We study $\Xi$-$^4{\mathrm{He}}$ ($\alpha$) momentum correlation functions in the high-energy nuclear collisions to investigate the nature of the $\Xi N$ interactions. 
	We employ the folding $\Xi \alpha$ potential based on the lattice QCD $\Xi N$ interactions to compute the correlation function. The $\Xi \alpha$ potential supports a Coulomb-assisted bound state ${}^5_{\Xi}\mathrm{H}$ in the $\Xi^-\alpha$ channel, while the $\Xi^0\alpha$ channel is unbound. To examine the sensitivity of the correlation function to the nature of the $\Xi \alpha$ interaction, we vary the potential strength simulating stronger and weaker interactions.
    The result of the correlation function is sensitive to the existence of the bound state in the $\Xi^0 \alpha$ channel, and the characteristic behavior of the bound state remains also in the $\Xi^-\alpha$ correlation with the Coulomb interaction.
    The effect of the repulsive core of the $\Xi\alpha$ potential can be found in the correlation from the small source as the distinctive dip in the intermediate momentum region.  
\end{abstract}
% 25.75.Gz : Particle correlation and fluctuations,
% 21.30.Fe : Forces in hadronic systems and effective interactions
% 13.75.Ev : Hyperon-nucleon interaction
\maketitle

\section{Introduction}\label{sec:Intro}
% alpha Xi system
%(About $\Xi\alpha$ system.. )

% Xi N (Lambda\Lambda) interactions
Understanding the properties of the Hyperon ($Y$)-Nucleon ($N$) interactions is a 
long standing problem in the hypernuclear physics. 
Among others, the $\Xi N$ interaction in the strangeness $S=-2$ sector is important to pin down the role of $\Xi$ in neutron stars and possible existence of the $H$ dibaryon~\cite{Jaffe:1976yi}. There are four independent components in the $s$-wave $\Xi N$ interaction, $ {}^{11}S_0,{}^{31}S_0,{}^{13}S_1,{}^{33}S_1$ where the notation $^{2I+1,2s+1}L_J$ is adopted with the spin $s$, isospin $I$, and the total angular momentum $J$.

% N Xi interactions : current status
Through the recent experimental observations of the $\Xi$ hypernuclei~\cite{Nakazawa:2015joa,J-PARCE07:2020xbm,Yoshimoto:2021ljs},
the  $\Xi N $ interaction is considered to be moderately attractive in total.
This is consistent with the recent theoretical analysis with the chiral effective field theories~\cite{Haidenbauer:2015zqb,Li:2018tbt,Haidenbauer:2018gvg} and the lattice QCD calculations~\cite{Sasaki:2017ysy,HALQCD:2019wsz}.
%Sasaki:2019qnh}.
On the other hand, the scattered results of the theoretical predication on the light $\Xi$ hypernuclei~\cite{Fujiwara:2006fj,Garcilazo:2016ylj,Filikhin:2017fog,Hiyama:2019kpw,Le:2021gxa,Miyagawa:2021krh,Friedman:2021rhu,Hiyama:2022jqh,Friedman:2022huy} indicate that the detailed understanding of the $\Xi N$ interactions is still insufficient and further constraints through the experimental observables are required.

% Femtoscopy 
In the past few years, 
the femtoscopic study using the momentum correlation function from the high-energy collisions has
provided new insights into hyperon interactions. 
The correlation function is known to be sensitive to the low-energy interaction and the particle emission source. 
In addition, the femtoscopy approach is advantageous for the multistrangeness sector which is not easily studied by the traditional scattering experiments.
For the baryon pairs including strangeness, $p\Lambda$~\cite{
STAR:2005rpl,ALICE:2018ysd,
%Adams:2005ws,Acharya:2018gyz,
ALICE:2021njx}, $p\Sigma^0$~\cite{ALICE:2019buq}, $p\Xi^-$~\cite{
ALICE:2019hdt,
%Acharya:2019sms,
ALICE:2020mfd,Isshiki:2021bqh} $\Lambda\Lambda$~\cite{
STAR:2014dcy,ALICE:2018ysd,ALICE:2019eol,
%Adamczyk:2014vca,Acharya:2018gyz,Acharya:2019yvb,
Isshiki:2021bqh},
${\Lambda\Xi}$~\cite{ALICE:2022uso}, $\Xi\Xi$~\cite{Isshiki:2021bqh}
 and $p\Omega$~\cite{
STAR:2018uho,
%Adams:STAR:2018uho,
 ALICE:2020mfd}  correlation data have been reported experimentally. 

% Xi N femtoscopy
For the study of the $\Xi N $ interactions,
$p\Xi^-$ correlation functions from the $pp$~\cite{ALICE:2020mfd} and $p$Pb collisions~\cite{
%Acharya:2019sms
ALICE:2019hdt} are measured by the ALICE collaboration.
The corresponding theoretical studies can be found in Refs.~\cite{Haidenbauer:2018jvl,Kamiya:2021hdb,Haidenbauer:2022suy,Haidenbauer:2022esw,Liu:2022nec}.
In Ref.~\cite{Kamiya:2021hdb}, the ALICE data are shown to be consistent with the theoretical correlation function calculated with the lattice QCD $\Xi N$-$\Lambda\Lambda$ coupled-channel potential~\cite{HALQCD:2019wsz}.
%Sasaki:2019qnh}. 
The measured correlation function is also in good agreement with the results in Refs.~\cite{Haidenbauer:2018jvl,Haidenbauer:2022suy,Haidenbauer:2022esw,Liu:2022nec}. The studies of Ref.~\cite{Liu:2022nec} are based on the covariant chiral effective field theory~\cite{Li:2018tbt} fitted to the phase shifts calucated by the HAL QCD potentials~\cite{Sasaki:2017ysy}.
References.~\cite{Haidenbauer:2018jvl,Haidenbauer:2022suy,Haidenbauer:2022esw} are based on the chiral effective field theory~\cite{Haidenbauer:2015zqb,Haidenbauer:2018gvg}.
The spin-isospin components of the $\Xi N$ interaction exhibit different properties from the lattice QCD potential. 
This is because the enhancement of the $\Xi N$ correlation is originated mainly in the strong attraction in the ${}^{11}S_0$ component, which shows similar behavior in both potentials.
Thus, with these studies, strongly attractive nature of the ${}^{11}S_0$ interaction is well confirmed, while the other components still lack the thorough understanding. 

% Xi nucleus femtoscopy
To impose further constraints on the $\Xi N $ interaction by the femtoscopic study, 
it is interesting to consider the correlation between $\Xi$ and nuclei. The correlation functions including nuclei have recently been studied theoretically~\cite{Mrowczynski:2019yrr,Bazak:2020wjn,Mrowczynski:2021bzy,Haidenbauer:2020uew,Ogata:2021mbo,Viviani:2023kxw,Jinno:2024tjh,Kohno:2024tkq}
and experimentally~\cite{Singh:2022qmg,Hu:2023iib}.
Because the $\Xi$-nucleus correlation picks up different combination of the spin-isospin components from the $p\Xi^-$ correlation function, further constraints on the $\Xi N$ interactions can be imposed by the study of the $\Xi$-nucleus correlation.
In this direction, the deuteron-$\Xi$ correlation is studied in Ref.~\cite{Ogata:2021mbo}, 
and it is found that the clear enhancement by the strong interaction appears in the correlation function, while the deuteron breakup effect is  sufficiently weak. However, in general, complicated three-body dynamics plays an important role in the correlation with the weakly bound deuteron~\cite{Viviani:2023kxw}.

% Xi alpha femtoscopy
Another possible candidate for the femtoscopy is the pair of a hyperon and $^4{\mathrm{He}}$ ($\alpha$) particle. The tightly bound nature of $\alpha$ justifies the 
two-body treatment of the hyperon-$\alpha$ system. In fact, a recent theoretical study of the $\Lambda \alpha$ correlation function shows that the femtoscopic analysis of  $\Lambda \alpha$ helps to investigate the $\Lambda N$ interaction~\cite{Jinno:2024tjh}. 
It is also found that the $\Lambda\alpha$ correlation function exhibits unique feature not seen in the hadron-hadron correlation functions, due to the relatively longer interaction range of the hyperon-nucleus interaction~\cite{Jinno:2024tjh}.
In addition, the $\Xi\alpha$ correlation can be used to disentangle the various spin-isospin components of the $\Xi N$ interaction. 
The $\Xi\alpha$ interaction is expressed by the weighted sum of the $\Xi N$ interaction components as 
$ \left[ V({}^{11}S_0)+3V({}^{31}S_0)+3V({}^{13}S_1)+9V({}^{33}S_1)\right]/16$.
Namely, the $\Xi\alpha$ interaction is dominated by the ${}^{33}S_1$ component of the $\Xi N$ interaction~\cite{Hiyama:2022jqh}.
Thus the $\Xi\alpha$ correlation brings the information complementary to the $p\Xi^-$ correlation. 

% bound state of ^5_Xi H 
There are two charge channels, $\Xi^0\alpha$ and $\Xi^-\alpha$; the former is governed solely by the strong interaction, while additional attraction is given to the latter by the Coulomb interaction. The theoretical studies of the $\Xi$ hypernuclei~\cite{Hiyama:2022jqh} predict the existence of a shallow bound state ${}^5_{\Xi}\mathrm{H}$ in the $\Xi^-\alpha$ channel but no bound state in the $\Xi^0\alpha$ channel. Such a shallow bound state that vanishes when the Coulomb interaction is switched off is called Coulomb-assisted bound state.
On the other hand, in Ref.~\cite{Le:2021gxa}, bound states are predicted in both charge channels due to the relatively stronger $\alpha\Xi$ interaction and the binding energy of ${}^5_{\Xi}\mathrm{H}$ is found to be larger than the result of Ref.~\cite{Hiyama:2022jqh}. 
The binding energy of ${}^5_{\Xi}\mathrm{H}$ depends on the employed $\Xi N$ interactions, mainly due to the difference in the ${}^{33}S_1$ component which gives the largest contribution~\cite{Le:2021gxa}.
Because it is known that the correlation function is sensitive to the existence and the binding energy of the bound state~\cite{Morita:2019rph,Kamiya:2015aea}, 
the $\Xi \alpha $ correlation function is expected to be helpful to elucidate the nature of the ${}^5_{\Xi}\mathrm{H}$ state and the $\Xi \alpha$ interaction. 
However, it is unknown how the correlation functions of the charged pair behave when the Coulomb-assisted bound state emerges.
In contrast with these, Ref.~\cite{Fujiwara:2006fj} reported that the $G$-matrix calculation based on the SU(6) quark-model baryon–baryon interaction does not support  a bound state.

%This study 
In this study, we give theoretical predictions for the $\Xi\alpha$ correlation function which can be measured in future experiments. 
We employ the $\Xi\alpha$  folding potential~\cite{Hiyama:2022jqh} constructed with the HAL QCD $\Xi N$ interactions~\cite{HALQCD:2019wsz}.
%Sasaki:2019qnh}. 
To consider the theoretical uncertainties, we vary the strength of the potential to realize the case with and without the $\Xi\alpha$ bound state and see how the correlation is affected by the variation of the potential strength. 
Furthermore, to investigate the relevance of the detailed shape of the interaction potential, 
we prepare the simple Gaussian potential models and compare the results with those by the folding potentials. 

% This article 
This article is organized as follows.
In Sec.~\ref{Sec:Formalism}, we briefly review the current status of the $\Xi\alpha$ interactions and introduce the framework to calculate the $\Xi\alpha$ correlation function. 
In Sec.~\ref{Sec:Results}, we show the results of the $\Xi\alpha$ correlation functions with examining various potentials and approximations. 
Section~\ref{Sec:Conclusion} is devoted to a summary and concluding remarks.

\section{formalism} \label{Sec:Formalism}

\begin{figure}[tbp]
	\begin{center}
		\includegraphics[width=0.45\textwidth]{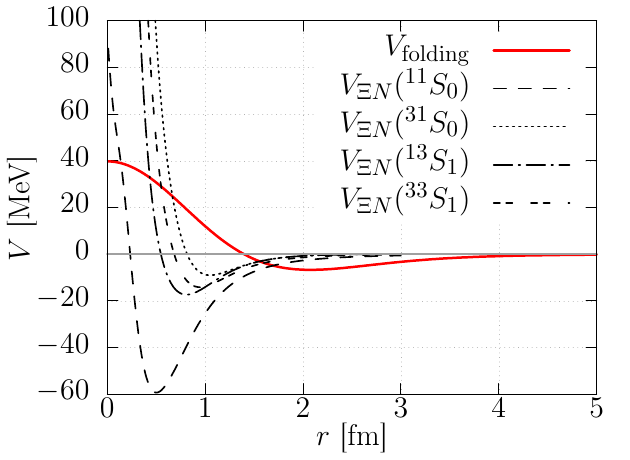}
		\caption{The $\Xi\alpha$ folding potential 
  $V_{\rm folding}$~\cite{Hiyama:2022jqh} and the HAL QCD $\Xi N$ potentials $V_{\Xi N}({}^{2I+1,2s+1}L_J)$~\cite{HALQCD:2019wsz}.
%Sasaki:2019qnh}. 
 For $V_{\Xi N}$, the central values of $t =12$ data are plotted.}
		\label{fig:V_Hiyama}
	\end{center}
\end{figure}

% Xi  \alpha potential 
In this study, we treat the $\Xi\alpha$ pair as a two-body system of $\Xi$ and $\alpha$, without considering the $\alpha$ breakup processes. We employ the $\Xi\alpha$ folding potential developed in Ref.~\cite{Hiyama:2022jqh}. The folding potential is constructed by using the HAL QCD $\Xi N$ potentials~\cite{HALQCD:2019wsz}
%Sasaki:2019qnh}.
where the small $\Xi N$-$\Lambda\Lambda$ coupling in the ${}^{11}S_0$ channel is renormalized into the effective single channel $\Xi N$ potential with a real coupling strength.
This folding $\Xi\alpha$ potential is parametrized by the sum of the Gaussians as 
\begin{align}
	V_{\rm folding}(r)= \sum_{i=1}^{20}V_i \exp(-\nu_i r^2), \label{eq:V_Hiyama}
\end{align}
where $V_i$ is the potential strength and $\nu_i$ is the Gaussian range. 
Because the tiny decay process $\Xi N\to \Lambda \Lambda$ is not included in the construction, $V_i$ are the real parameters.
The shape of the $\Xi\alpha$ folding potential is shown in Fig.~\ref{fig:V_Hiyama} together with the HAL QCD $\Xi N$ potentials. 
It is seen that the potential $V_{\rm folding}$ has a repulsive core at short distance and an attractive pocket in the longer range, as in the $\Xi N$ interactions.
Due to the spatial extent of the $\alpha$ particle, both the central repulsion and the attractive pocket are smeared from the $\Xi N$ interactions. As a consequence, the  range of the $\Xi \alpha$ potential is longer than the $\Xi N$ interactions.
%The attraction of $V_{\rm folding}$ is not strong enough to support a bound state.
By solving the Schr\"odinger equation, 
the attraction of $V_{\rm folding}$ turns out to be not strong enough to support a bound state.

In the practical calculations, we need to consider the $\Xi^0\alpha$ and $\Xi^-\alpha$ systems separately, due to the Coulomb interaction. 
The $\Xi^0\alpha$ system, where only the strong interaction works, does not have a bound state.
The scattering length $a_0$ and the effective range $r_e$ are summarized in Table~\ref{tab:a0_alphaXi}. 
Note that we employ the nuclear physics convention for the scattering length where positive (negative) $a_0$ corresponds to the repulsive interaction or the strongly attractive interaction with a bound state (weakly attractive interaction without a bound state).
This significantly large scattering length $|a_0|\approx 500$ fm implies that the system is very close to the unitary limit~\cite{Braaten:2004rn} and the bound state would appear if the attraction were slightly stronger.
The larger effective range $r_e$ than usual hadron potentials reflects the longer interaction range of the $\Xi\alpha$ potential~\eqref{eq:V_Hiyama}.

% alpha Xi^- system and Coulomb intneraction
For the $\Xi^-\alpha$ system, the further attraction by the Coulomb interaction acts as 
\begin{align}
	V_\mathrm{Coulomb} (r)= -\frac{2\alpha}{r}, \label{eq:V_C}
\end{align}
with the fine-structure constant $\alpha$\footnote{In this study, we use the point-like Coulomb potential neglecting the finite spatial distribution of the wave function of the $\alpha$ particle.  This is because the wave function for the point-like Coulomb potenital is analytically known and it is much easier to numerically calculate the correlation function. In fact, the folding effect of Coulomb interaction changes the binding energy of ${}^5_{\Xi}\mathrm{H}$ just by $\approx 0.02$ MeV.}.
By solving the Schr\"odinger equation with $V=V_{\rm folding}+V_\mathrm{Coulomb}$, we find  a Coulomb-assisted shallow bound state ${}^5_{\Xi}\mathrm{H}$ with the binding energy
\begin{align}
	B=0.47 \physdim{MeV} .
\end{align}
Note that this is distinguishable from the Coulomb bound states, which emerge from the purely Coulombic attraction.
The binding energy of the ground state by the pure Coulomb potential is given as 
\begin{align}
	B^{\mathrm{Coulomb}} = \frac{\alpha}{ a^{\mathrm{Bohr}}} =  0.104\physdim{MeV}, \label{eq:B_Coulomb}
\end{align}
with the Bohr radius of the $\Xi^-\alpha$ system $a^{\mathrm{Bohr}} = (2\mu_{\Xi^-\alpha} \alpha)^{-1}= 13.9$ fm.  
Compared with $B^{\mathrm{Coulomb}} $,  the binding energy $B=0.47$ MeV is about 0.3 MeV deeper and its dominant contribution can be regarded as originating from the strong interaction $V_{\rm folding}$.

\begin{figure}[tbp]
	\begin{center}
		\includegraphics[width=0.45\textwidth]{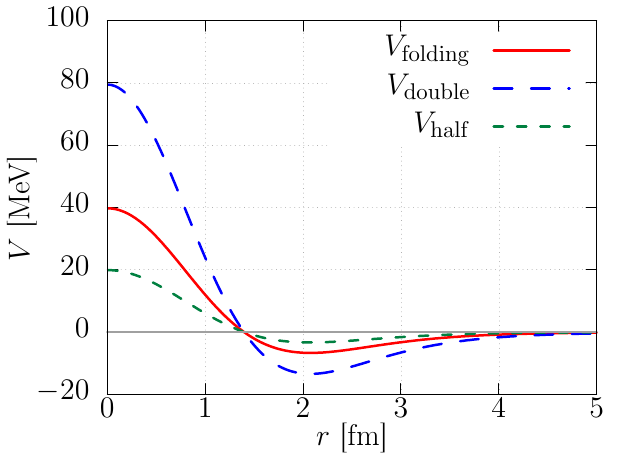}
		\caption{Comparison of the $\Xi\alpha$ folding potential $V_\mathrm{folding}$ and its variations of $V_{\rm double}$ and $V_\mathrm{half}$.}
		\label{fig:V_var}
	\end{center}
\end{figure}

% Variation of alpha Xi potential 
In the study of chiral effective field theory~\cite{Le:2021gxa}, the ${}^5_{\Xi}\mathrm{H}$ bound state with $B= 2.16$ MeV is found by using the  chiral next-to-leading order (NLO) $\Xi N$ interactions. This indicates that the chiral $\Xi\alpha$ interaction is more attractive than the HAL QCD based potential $V_{\rm folding}$.
On the other hand, because the strength of the $\Xi \alpha$ potential is to some extent ambiguous, the $\Xi\alpha$ potential might be less attractive than $V_{\rm folding}$ without generating the bound ${}^5_{\Xi}\mathrm{H}$. 
To examine the theoretical uncertainty of the $\Xi\alpha$ potential, we consider two variations of the potential; $V_{\rm double}   = 2V_{\rm folding}$ and  $V_{\rm half}=V_{\rm folding}/2$ for the stronger and weaker potentials, respectively.
These variations are shown in Fig.~\ref{fig:V_var}.
With the stronger potential $V_{\rm double}$, the ${}^5_{\Xi}\mathrm{H}$ binding energy is found to be 2.08 MeV, which is close to the value in the chiral NLO analysis. 
In addition, this stronger potential  $V_{\rm double}$ makes a bound state also in the $\Xi^0\alpha$ channel without the Coulomb attraction. Accordingly, the scattering length $a_0$ becomes positive. 
On the other hand, the binding energy of ${}^5_{\Xi}\mathrm{H}$ with the weaker potential $V_{\rm half}$ is 0.18 MeV, which is of the same order as the binding energy of the Coulomb bound state~\eqref{eq:B_Coulomb}. 
As expected, there is no bound state in the $\Xi^0\alpha$ system with $V_{\rm half}$. 
The scattering lengths $a_0$, effective ranges $r_e$, and the binding energies of the $\Xi\alpha$ systems for different potentials are summarized in Table~\ref{tab:a0_alphaXi} and \ref{tab:bd_alphaXi}.

\begin{table}
        \caption{The scattering length $a_0$ and the effective range $r_e$ of the $\Xi^0\alpha$ scattering calculated with the folding potential $V_{\rm folding}$ and its variations $V_{\rm double} $ and $V_{\rm half}$.}
		\label{tab:a0_alphaXi}
\begin{ruledtabular}
		\begin{tabular}{ccc}
			potential& $a_0$ [fm] & $r_e$ [fm]\\
			\hline
			$V_{\rm folding}$ & $-522.8$ & $4.50$ \\
			$V_{\rm double} = 2V_{\rm folding}$ & $\phantom{-}6.39$   & $3.01$ \\
			$V_{\rm half} = V_{\rm folding}/2$ & $-3.39$  &  7.36 \\
		\end{tabular}
\end{ruledtabular}
\end{table}

% correlation function 
In the present framework, the $\Xi\alpha$ system is treated as a single-channel scattering. For a given $\Xi\alpha$ potential, the
momentum correlation function $C(\bm{q})$ in the high-energy nuclear collisions can be calculated by the Koonin-Pratt (KP) formula~\cite{Koonin:1977fh,Pratt:1986cc,Pratt:1990zq,ExHIC:2017smd};
\begin{align}
	C(\bm{q}) = \int d^3 r S(\bm{r})\left|\Psi^{(-)}(\bm{q};\bm{r})\right|^2, \label{eq:KP}
\end{align}
where $\bm{q}$ is the relative momentum in the pair rest frame, $S(\bm{r})$ is the normalized source function, and $\Psi^{(-)}(\bm{q};\bm{r})$ is the relative wave function with the outgoing boundary condition, calculated by the potential. In this study, we employ the static Gaussian source function $S_R(r)\equiv \exp(-r^2/4R^2)/(4\pi R^2)^{3/2}$ with the source size\footnote{We note that $S(r)$ is the relative source function representing the distribution of the $\Xi\alpha$ pair emitted with the relative distance $r$. The relative source size is defined as $R=\sqrt{(R_\Xi^2+R_\alpha^2)/2}$ with $R_\Xi$ ($R_{\alpha}$) being the source size of the single $\Xi$ ($\alpha$) production. With this definition, the relative source is given by the gaussian with radius $\sqrt{2}R$~\cite{ExHIC:2017smd}.} $R$.
For the $\Xi^0\alpha$ pair, we consider the modification of the wave function from the plane wave only in the $s$-wave component, which is significantly distorted by the strong interaction in the low-momentum region.
On the other hand, for the $\Xi^-\alpha$ pair, because of the long range nature of the Coulomb interaction, one must use the Coulomb wave functions as the asymptotic form in all the partial waves, and introduce the strong interaction effect in the $s$ wave~\cite{Kamiya:2019uiw}.

\begin{table}
        \caption{The binding energy of the $\Xi^-\alpha$ and $\Xi^0\alpha$ systems. The Coulomb interaction is included in the $\Xi^-\alpha$ calculation. }
		\label{tab:bd_alphaXi}
\begin{ruledtabular}
		\begin{tabular}{ccc}
			potential&$\Xi^-\alpha$  [MeV] & $\Xi^0 \alpha$ [MeV]\\
			\hline
			$V_{\rm folding}$ & $0.47$  & - \\
			$V_{\rm double} $ & $2.08$   &  1.15 \\
			$V_{\rm half} $ & $0.18$  & -- \\
		\end{tabular}
\end{ruledtabular}
		%\caption{The $\Xi^-\alpha$ and $\Xi^0\alpha$ binding energy. The Coulomb interaction is included for the $\Xi^-\alpha$ calculation. }
\end{table}

\section{Results}\label{Sec:Results}

\begin{figure*}[tbp]
	\begin{center}
		\begin{minipage}{1\hsize}
			\includegraphics[width=0.33\textwidth]{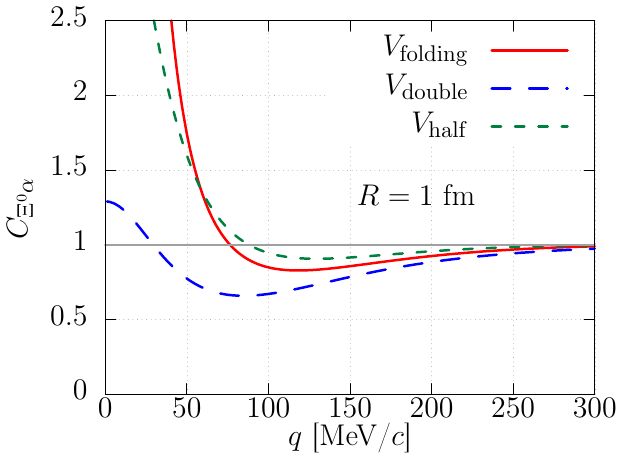}
			\includegraphics[width=0.33\textwidth]{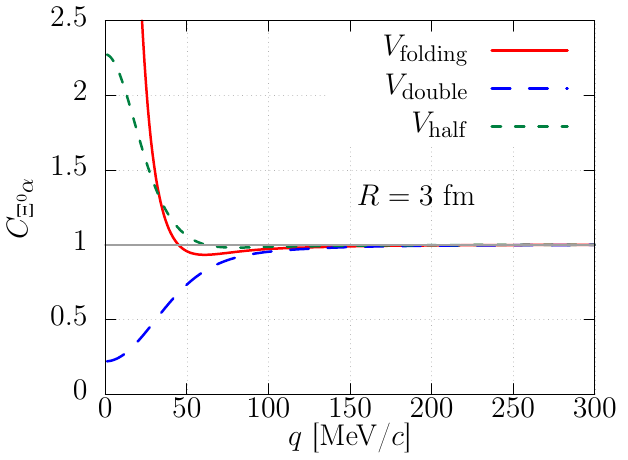}
			\includegraphics[width=0.33\textwidth]{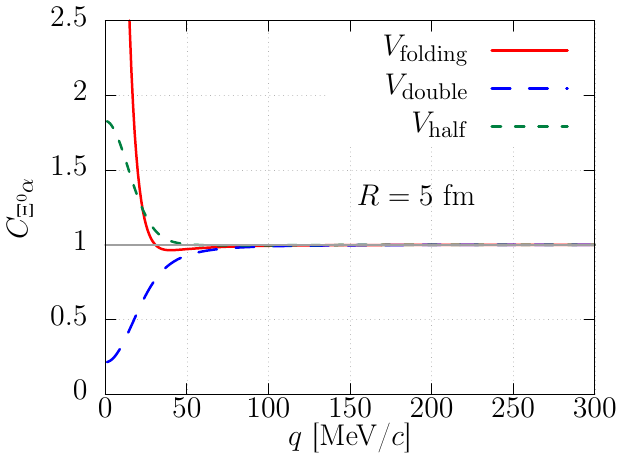}
		\end{minipage}
		\caption{The $\Xi^0\alpha$ correlation function with $R = 1$, $3$, and $5 $ fm. The results with $V_{\rm folding}$, $V_{\rm double}$, and $V_{\rm half}$ are shown by solid, long-dashed, and dashed lines respectively. }
		\label{fig:corr_Xi0}
	\end{center}
\end{figure*}

% Xi0 alpha correlation
First we discuss the $\Xi^0\alpha $ correlation function without the Coulomb attraction.  
The results with $V_{\rm folding}$, $V_{\rm double}$, and $V_{\rm half}$ for $R = 1, 3$, and $5$ fm are shown in Fig.~\ref{fig:corr_Xi0}.\footnote{The source size $R=1$ fm might look small for the emission of the $\alpha$ particle with the charge radius $\sim 1.68$ fm~\cite{Krauth:2021foz}. As we mentioned, however, the relative source function $S(r)$ has the gaussian width $\sqrt{2}R$, and the probability distribution of the relative distance $r$ is given by $4\pi r^2 S(r)$~\cite{Mihaylov:2018rva}. As a consequence, with $R=1$ fm, the mean distance between the emitted pair is about $\langle r\rangle =4R/\sqrt{\pi}\sim 2.26$ fm.}
We see that $C_{\Xi\alpha}$ with $V_{\rm double}$ shows qualitatively different behavior from those with $V_{\rm folding}$ and $V_{\rm half}$. This is attributed to the existence of the bound state by $V_{\rm double}$. In addition, the very strong enhancement of the correlation with $V_{\rm folding}$ at small momentum reflects the significantly large scattering length $|a_0|>500$ fm. As a consequence,  three different potential models adopted here are distinguishable by the measurement of the $\Xi^0\alpha$ correlation function, in particular for the large source $R=3$-5 fm.
We find that the result with $V_{\rm double}$ shows the suppression or bump structure depending on the source size $R$.
This is a typical feature of $C_{\Xi \alpha}$ for the attractive interaction with a bound state.
On the other hand, the correlation functions with $V_{\rm folding}$ and $V_{\rm half}$ show the enhancement in the low momentum region characteristic for an attractive interaction without a bound state, but a
dip structure in the intermediate momentum region ($q \sim 200$ MeV/$c$) is found. 
The dip structure is more prominent in $C_{\Xi \alpha}$ with a small source, $R=1$ fm.
Because such dip structure is not seen in the model calculation with simple attraction~\cite{Morita:2016auo}, this should be related to the detailed shape of the $\Xi \alpha$ potential.

\begin{figure*}[tbp]
	\begin{center}
		\includegraphics[width=0.32\textwidth]{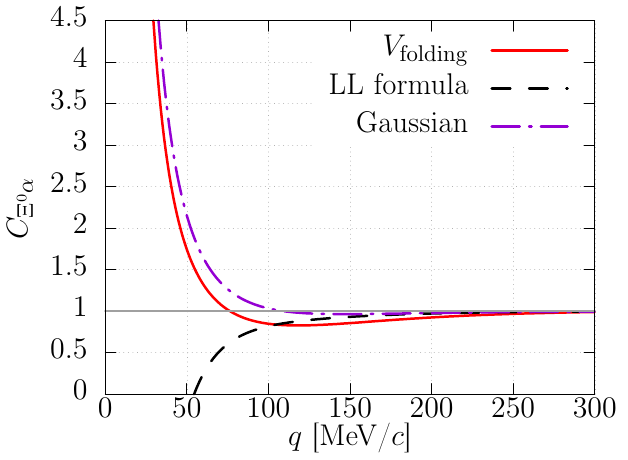}
		\includegraphics[width=0.32\textwidth]{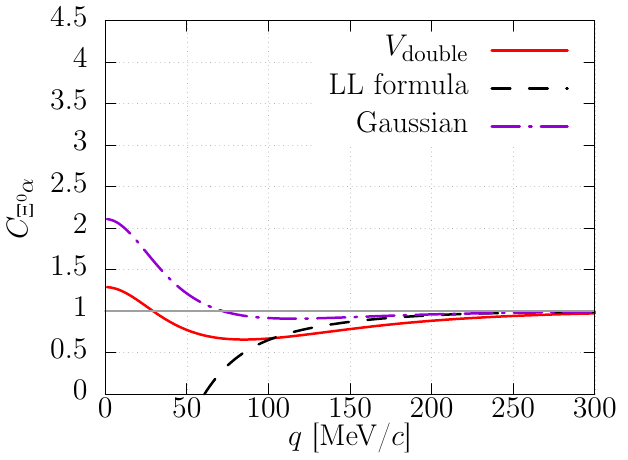}
		\includegraphics[width=0.32\textwidth]{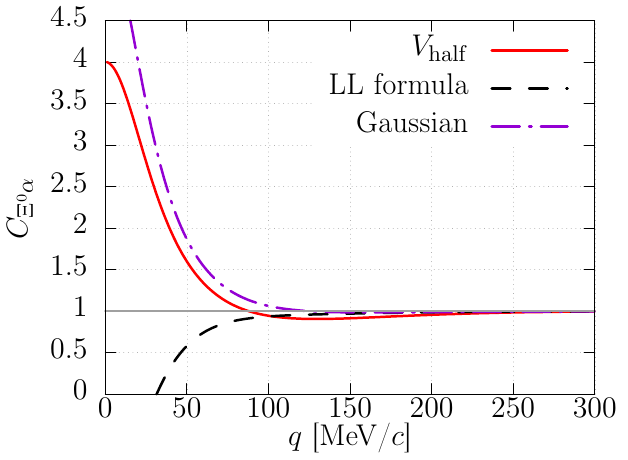}
		\caption{Comparison of $\Xi^0\alpha$ correlation functions for $R=1$ fm with the KP formula for the folding potential (solid lines), the Gaussian potential model (dash-dotted lines), and the LL formula estimation (dashed lines).  The left, central, and right panel show the results with $V_{\rm folding}$, $V_{\rm double}$, and $V_{\rm half}$, respectively.} 
		\label{fig:corr_LL}
	\end{center}
\end{figure*}

% Gaussian
To see the effect of the shape of the $\Xi\alpha$ folding potential, we introduce the purely attractive one range Gaussian potential given as 
\begin{align}
	V_{\mathrm{Gaussian}}(r) = V_0 \exp(-r^2/b^2),
\end{align}
with the potential strength $V_0$ and the range parameter $b$. 
We construct the Gaussian potentials by choosing the range parameter as $b=3$ fm and tuning $V_0$ to reproduce the scattering length $a_0$ in Table~\ref{tab:a0_alphaXi} for each potential. We have checked that the qualitative conclusions given below remain unchanged under the variation of the value of $b$.
The  correlation functions by the Gaussian potentials with $R=1$ fm are compared with the results from the original folding potentials $V_{\rm folding}$, $V_{\rm double}$, and $V_{\rm half}$ in Fig.~\ref{fig:corr_LL}. 
We find that the Gaussian potentials qualitatively reproduce the results of the original folding potentials, while the correlation in the small momentum region is somehow overestimated. In particular, the Gaussian potentials corresponding to $V_{\rm folding}$ and $V_{\rm half}$ without a bound state provide the enhancement of the correlation without a dip in the intermediate momentum region, as expected. 
In other words, the folding potentials with a repulsive core give the suppression of the correlation functions in this region, causing a dip structure.
%This suppression can be understood as the effect of the repulsive core. 
Thus, we conclude that the characteristic suppression found in the $\Xi^0\alpha$ correlation with $R=1$ fm in the intermediate momentum region is caused by the repulsive core of the folding potential. 
This means that the correlation function from the small source may be useful to investigate the existence and strength of the repulsive core of the $\Xi\alpha$ interaction. 

\begin{figure}[tbp]
	\begin{center}
		\includegraphics[width=0.32\textwidth]{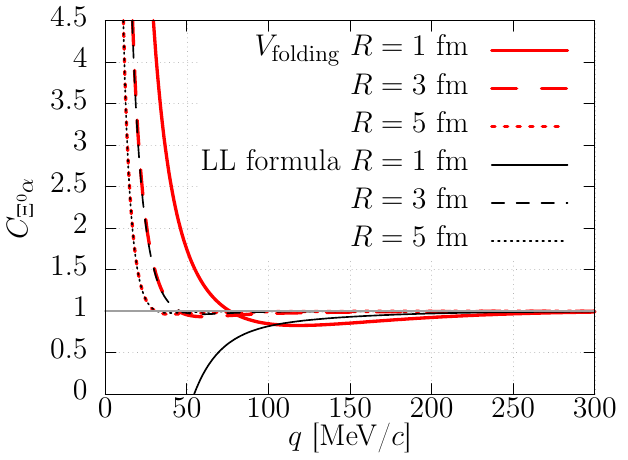}
		\caption{Comparison of $\Xi^0\alpha$ correlation functions by $V_{\rm folding}$ with the KP formula (thick lines) and the LL formula estimation (thin lines) for different source sizes. The results with $R=1$, 3, and 5 fm are denoted by solid, dashed, and dotted lines, respectively.}
		\label{fig:corr_LL_R}
	\end{center}
\end{figure}

% LL formula
To further discuss the effect of the potential shape to the correlation, we evaluate the correlation functions with 
the Lednicky-Lyuboshits (LL) formula~\cite{Lednicky:1981su,ExHIC:2017smd} 
\begin{align}
	C_{\mathrm{LL}}(q)=&1 + 	\frac{|f(q)|^2}{2R^2}F_3\left(\frac{r_e}{R}\right) \notag \\
	 & +\frac{2\mathrm{Re} f(q)}{\sqrt{\pi}R}F_1(2qR)
	 -\frac{\mathrm{Im} f(q)}{R} F_2(2qR), \label{Eq:LL}
\end{align}
where $F_1(x)=\int_0^x dt\,e^{t^2-x^2}/x$, $F_2(x)=(1-e^{-x^2})/x$, $F_3(x)=1-x/2\sqrt{\pi}$,
and $f(q)= 1/ (-1/a_0+ r_e/2 q^2 -iq)$ is the $s$-wave $\Xi\alpha$ scattering amplitude calculated by the effective range expansion with the threshold parameters in Table~\ref{tab:a0_alphaXi}.
The LL formula is obtained from the KP formula by approximating the full wave function by the asymptotic wave function.
In Fig.~\ref{fig:corr_LL}, we compare the results by the KP formula~\eqref{eq:KP} with the corresponding ones by the LL formula for $V_{\rm folding}$, $V_{\rm double}$, and $V_{\rm half}$ potentials with the source size $R=1$ fm.
As shown in Fig.~\ref{fig:corr_LL}, for $R= 1$ fm case, the results with the LL formula do not reproduce those with the KP formula for all potentials. 
On the other hand, as shown in Fig.~\ref{fig:corr_LL_R} for $R=3$ and 5 fm cases, the LL formula gives the good approximation of the KP formula results.
This failure of the LL formula for small source also indicates that the detailed shape of the potential affects the correlation function, which is qualitatively consistent with what is found in the study of the $\Lambda\alpha$ correlation function~\cite{Jinno:2024tjh}.

\begin{figure*}[htbp]
	\begin{center}
			\begin{minipage}{1\hsize}
		\includegraphics[width=0.33\textwidth]{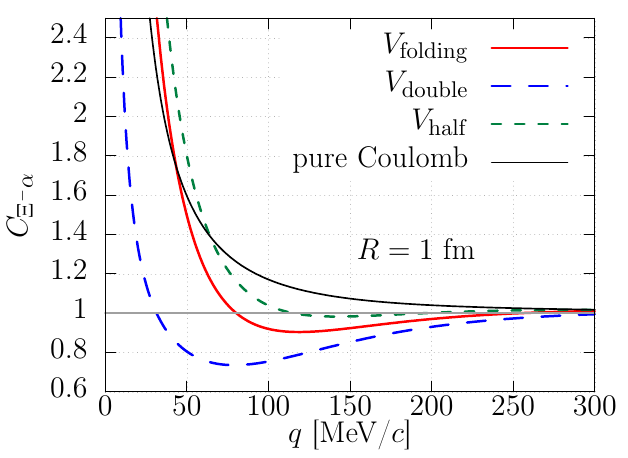}
		\includegraphics[width=0.33\textwidth]{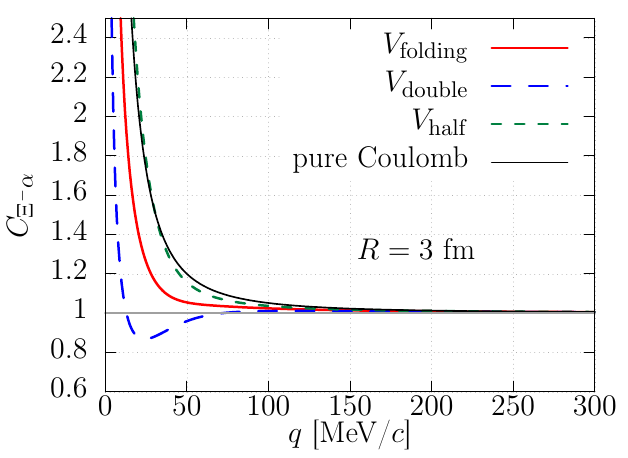}
		\includegraphics[width=0.33\textwidth]{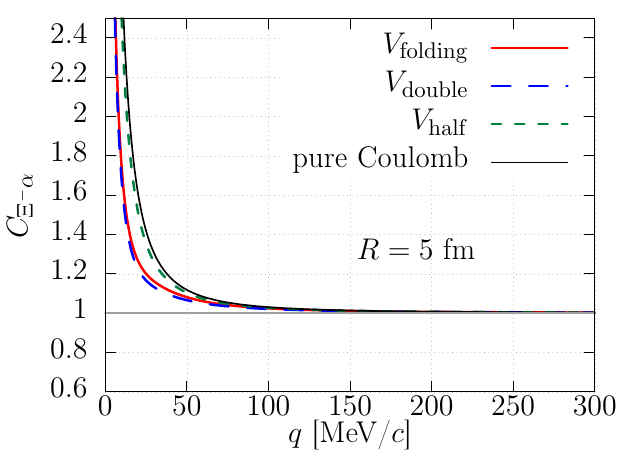}
		\end{minipage}
		\caption{The $\Xi^-\alpha$ correlation function with $R=1, 3 $ and $5$ fm. The results with $V_{\rm folding}$, $V_{\rm double}$, and $V_{\rm half}$ are shown by solid, long-dashed, and dashed lines respectively. The pure Coulomb result, where the strong interaction is switched off,  is shown by the solid thin line for comparison.}
		\label{fig:corr_Xim}
	\end{center}
\end{figure*}

% Xi- alpha
Finally, we show the results of the $\Xi^-\alpha$ correlation functions in Fig.~\ref{fig:corr_Xim}. 
Due to the Coulomb attraction, $C_{\Xi\alpha}$ shows strong enhancement at the low momentum for all potentials.
The effect of the strong interaction emerges as the deviation from the pure Coulomb result, where the strong interaction is switched off.
In contrast with the $\Xi^0\alpha$ correlation, the difference between the adopted potentials in larger source is smeared by the Coulomb attraction. Nevertheless, with a good resolution of measurement, it may be possible to distinguish different potentials by the correlation function with $R=1$-3 fm. 
Through the comparison with $C_{\Xi^0\alpha}$ in Fig.~\ref{fig:corr_Xi0}, we find that the results with $V_{\rm double}$ and $V_{\rm half}$  in the low momentum region is simply enhanced due to Coulomb force from $C_{\Xi^0\alpha}$.
On the other hand, $C_{\Xi^-\alpha}$ with $V_{\rm folding}$ with $R=3$ and 5 fm is smaller than the pure Coulomb case while $C_{\Xi^0\alpha}$ shows the enhancement. 
Namely, the correlation function of $V_{\rm folding}$ shows enhancement for the small source and suppression for the large source with respect to the pure Coulomb result.
This is nothing but the source-size dependence of the correlation function with a shallow bound state. 
This means that, when the Coulomb-assisted bound state exists,  the typical source-size dependence can be observed in the $\Xi^-\alpha$ correlation function as the difference from the pure Coulomb result.

\section{Conclusion}\label{Sec:Conclusion}

% result of Xi^0 alpha correlation
Toward elucidating  the $\Xi N $ interaction, we have discussed the $\Xi\alpha$ correlation function with the folding potential $V_{\rm folding}$~\cite{Hiyama:2022jqh} obtained with the latest $\Xi N$ lattice QCD interactions~\cite{HALQCD:2019wsz}, which give the large scattering length without a bound state (a shallow Coulomb-assisted bound state)
in the $\Xi^0\alpha$ ($\Xi^-\alpha$) channel. Two variants of the $\Xi\alpha$ potential, $V_{\rm double}$ and $V_{\rm half}$, are used to discuss the dependence on the binding energy and existence of the bound states for the correlation function. 
The $\Xi^0\alpha$ correlations with $V_{\rm double}$ for the source sizes $R=1,3$, and 5 fm show the source-size dependence typical to the system with a bound state. 
On the other hand, the correlations with $V_{\rm folding}$ and $V_{\rm half}$ exhibit the characteristic suppression in the intermediate momentum region on top of the strong enhancement typical to the system with attractive interactions without a bound state. 

% detailed shape effect 
To discuss the effect of the shape  of the $\Xi\alpha$ potential on the correlation function, 
the results with the folding potentials are compared with those by the purely attractive Gaussian potential model and of the LL formula. 
Because the dip structure in the intermediate momentum region is not seen in the Gaussian potential result, we conclude that
the central repulsion of the folding potential is the origin of the suppression causing the dip. 
Thus, the strength of the central repulsion of the $\Xi\alpha$ potential is reflected in the dip structure found in the intermediate momentum region. 
Since the correlation lineshape is affected by various factors, it is difficult to extract such detailed features of the potential directly from experimental data. Nevertheless, it is an important point to note here that, in the hyperon-$\alpha$ correlation, the repulsive core can give rise to a sufficiently visible dip structure in the correlation.

% $\Xi^-\alpha$ correlation 
The $\Xi^-\alpha$ correlation with $V_{\rm folding}$ shows the source-size dependence reflecting the emergence of the ${}^5_\Xi \mathrm{H}$ state.
This result ensures that the strength of the  $\Xi\alpha$ attraction, which originates in  the $\Xi N$ interaction, and the existence of the Coulomb-assisted bound state of ${}^5_\Xi \mathrm{H}$ can be investigated by the source-size dependence of the $\Xi^-\alpha$ correlation.

% small source function 
In this study, we take the source sizes $R=$ 1, 3, and 5 fm for the theoretical prediction for the small to large source experiments.  
These values correspond to the typical sizes used in the analysis of the two-hadron correlation functions obtained in the high-energy $pp$ and heavy ion collisions. 
However, the effective source size of the composite particle $\alpha$, which can also be formed by the coalescence of four nucleons emitted from the fireball, must be larger than those with single hadron emissions~\cite{Mrowczynski:2019yrr,Bazak:2020wjn,Mrowczynski:2021bzy}.
In fact, the $\Xi\alpha$ pair is treated as a pair of the point-like particles in the Koonin-Pratt formula~\eqref{eq:KP}, which gives the normalization of  the emitting source function.   
For a more realistic study of the $\Xi\alpha$ correlation, the five body scattering problem of four nucleons and alpha should be solved to take into account the coalescence effect, which is left as future work. 

We have shown that the difference of the correlation functions from different $\Xi\alpha$ potentials, $V_{\rm folding}$, $V_{\rm double}$, and $V_{\rm half}$ can be well distinguished by the measurement of the $\Xi\alpha$ correlation functions, in particular by those from $R\sim 3$ fm source. 
%Future experiments
To measure the $\Xi\alpha$ correlation in experiments, it is desirable to use the high-energy collisions with the central energy $\sqrt{s}_{NN}  <  10 $ GeV, where a bunch of $\alpha$ is produced according to the estimation in Ref.~\cite{Andronic:2010qu}. 
For this reason, it is anticipated that the $\Xi\alpha$ correlation can be experimentally measured in the facilities such as FAIR~\cite{Ozawa:2022sam}, NICA, and J-PARC-HI~\cite{CBM:2016kpk} to elucidate the $\Xi\alpha$ interaction and its origin of the $\Xi N$ interactions.

\begin{acknowledgments}
We thank Emiko Hiyama for providing  the $\Xi\alpha$ folding potential and for helpful support. 
We also thank Kouichi Hagino, Yudai Ichikawa, Johann Haidenbauer, Andreas Nogga, Hoai Le, and Avraham Gal for useful discussions and comments.
This work was supported in part by the 
Grants-in-Aid for Scientific Research from JSPSs 
(Grants 
No.~JP23H05439, and % Kiban S (Hyodo)
No.~JP22K03637), % Kiban C (Hyodo)
by 
JST, the establishment of university fellowships toward
the creation of science technology innovation,
Grant No.~JPMJFS2123, 
by JST SPRING, Grant No. JPMJSP2110, 
and
by the Deutsche Forschungsgemeinschaft (DFG) and the National Natural Science Foundation of China (NSFC) through the funds provided to the Sino-German Collaborative Research Center ``Symmetries and the Emergence of Structure in QCD" (NSFC Grant No. 12070131001, DFG Project-ID 196253076 -- TRR 110).	
\end{acknowledgments}

\end{document}